\documentclass[aps,prb,floats,twocolumn,10pt]{revtex4-1}
\usepackage{graphicx}
\usepackage{amssymb}
\usepackage{amsmath}
\newcommand\numberthis{\addtocounter{equation}{1}\tag{\theequation}}
\usepackage{color}
\usepackage[dvipsnames]{xcolor}
\usepackage{slantsc}
\usepackage{float}
\graphicspath{{./FIG_FINALI/}}
\usepackage{bbold}
\usepackage[caption=false]{subfig}
\usepackage{array}
\usepackage{braket}
\usepackage{scrextend}
\usepackage{hyperref}
\usepackage[normalem]{ulem}
\begin{document}
\title{Spin-phonon coupling parameters from maximally localized Wannier functions and first principles electronic structure: the case of durene single
crystal}
\author{Subhayan Roychoudhury and Stefano Sanvito}
\affiliation{School of Physics and CRANN Institute, Trinity College, Dublin 2, Ireland}

\begin{abstract}
Spin-orbit interaction is an important vehicle for spin relaxation. At finite temperature lattice vibrations modulate the spin-orbit interaction and
thus generate a mechanism for spin-phonon coupling, which needs to be incorporated in any quantitative analysis of spin transport. Starting 
from a density functional theory \textit{ab initio} electronic structure, we calculate spin-phonon matrix elements over the basis of maximally 
localized Wannier functions. Such coupling terms form an effective Hamiltonian to be used to extract thermodynamic quantities, within a 
multiscale approach particularly suitable for organic crystals. The symmetry of the various matrix elements are analyzed by using the
$\Gamma$-point phonon modes of a one-dimensional chain of Pb atoms. Then the method is employed to extract the spin-phonon coupling 
of solid durene, a high-mobility crystal organic semiconducting. Owing to the small masses of carbon and hydrogen spin-orbit is weak in durene
and so is the spin-phonon coupling. Most importantly we demonstrate that the largest contribution to the spin-phonon interaction originates from
Holstein-like phonons, namely from internal molecular vibrations.
\end{abstract}

\maketitle

\section{Introduction}
In a non-magnetic material the electrical resistance experienced by a charge carrier is independent of its spin. In contrast, when the material is magnetic 
the resistance typically depends on the relative orientation of the carrier spin and the local magnetization~\cite{Mott699}. This observation inspired the 
advent of the field of spin-electronics or spintronics~\cite{Wolf1488}, which concerns the injection, manipulation and detection of spins in a solid-state 
environment. A prototype spintronics device, the spin-valve~\cite{PhysRevB.43.1297}, consists of two ferromagnetic layers sandwiching a non-magnetic 
spacer~\cite{Sanvito2011}, which can display a metallic~\cite{PhysRevLett.61.2472,PhysRevB.39.4828}, insulating~\cite{PhysRevLett.74.3273} or 
semiconducting~\cite{FabianZutic,Awschalom} electronic structure. The carriers, which are spin-polarized by one ferromagnet, travel through the spacer to 
the other ferromagnet. If the spin direction is maintained during such transfer, then the total resistance of the device will depend on the mutual orientation
of the magnetization vectors of the two ferromagnets. It is then crucial to understand how the spin direction evolves during the motion of the carriers
through the spacer, and in particular to understand how this is preserved.

There are multiple possible sources of spin relaxation in a material, such as the presence of impurities, hyperfine 
interaction and spin-orbit (SO) coupling. A theoretical description of all such phenomena is needed for an accurate 
evaluation of the quantities related to spin relaxation. The relative dominance of one interaction over the others is 
typically highly dependent on the specific material. In this work, we shall focus on SO interaction, more specifically 
on the modulation of such interaction due to lattice vibrations. The spin of an electron interacts with the magnetic field 
generated by the relative motion of the nucleus about the electron, giving rise to SO interaction. At finite temperature the 
atoms of a solid vibrate with respect to their equilibrium positions with the amplitudes of such vibrations increasing with 
temperature. Such vibrations, the phonons, change the potential felt by the electrons, including the component due to SO 
coupling~\cite{PhysRevLett.63.442}. This effectively generates a mechanism for spin-phonon coupling~\cite{PhysRevB.86.245411},
which is key for the calculation of quantities related to spin-relaxation in many systems. It must be noted that in current literature 
the term `spin-phonon' coupling has been used to denote different effects. For instance in the study of multiferroic compounds
`spin-phonon coupling' indicates the modulation of the phonon frequencies due to changes in the magnetic 
ordering.~\cite{PhysRevB.88.094103,PhysRevB.84.104440,PhysRevB.85.054417,BIROL2012227, PhysRevB.84.144409}
Here we are interested in the opposite, namely in the change of electronic structure brought by the vibrations, in particular
for the case of organic crystals.

Recent years have witnessed a growing interest in exploring the possibility of using organic crystal semiconductors for electronic 
and spintronic applications~\cite{doi:10.1063/1.1995748,ref1_nat,Forrest2004,C1CS15047B,8007720}. This stems from the high 
degree of mechanical flexibility, the light weight and the ease of synthesis and patterning that characterize organic compounds. 
In these systems covalently bonded organic molecules are held together by weak van der Waals interactions. Due to the weak bonds 
between the individual molecules, vibrational motions are prominent in organic crystals and the coupling of the vibrations to the 
charge carriers plays a crucial role~\cite{C0CS00198H} in the transport properties of such materials.

The presence of experimental evidence in support of different transport 
regimes~\cite{ADMA:ADMA99,doi:10.1063/1.1704874,doi:10.1063/1.2193801, PhysRevLett.97.256603,SakanoueTomoHenning} has 
generated a significant debate on whether the transport in organic crystals is dominated by delocalized band-like transport, 
as in covalently bonded inorganic semiconductors, by localized hopping, or by a combination of both. This can very well 
depend on the specific crystal and the experimental conditions, such as the temperature. Typically, in organic crystals the vibrational 
degrees of freedom are thought to introduce significant dynamical disorder~\cite{doi:10.1021/jp055432g} and thereby have paramount 
influence on the transport properties. Since the typical energies associated to lattice vibrations in organics are of the same order of 
magnitude of the electronic bandwidth, the coupling between carriers and phonons can not be treated by perturbation theory. Thus, in general, 
formulating a complete theoretical framework for the description of transport in organic crystals is more challenging than that for covalently 
bonded inorganic semiconductors~\cite{C0CS00198H,PhysRevB.69.075212}. Even more complex is the situation concerning spin transport, 
for which the theoretical description often relies on parameters extracted from experiments~\cite{PhysRevLett.102.156604}, or on 
approximate spin Hamiltonians~\cite{Sandip}

One viable option towards a complete \textit{ab initio} description of spin transport consists in constructing a multiscale approach, where 
information about the electronic and vibrational properties calculated with first-principles techniques are mapped onto an effective Hamiltonian 
retaining only the relevant degrees of freedom. For instance this is the strategy for constructing effective giant-spin Hamiltonians with spin-phonon
coupling for the study of spin relaxation in molecular magnets~\cite{Alessandro1,Alessandro2}. The approach presented here instead consists in
projecting the electronic structure over appropriately chosen maximally localized Wannier functions (MLWFs)~\cite{RevModPhys.84.1419}, 
which effectively define a tight-binding (TB) Hamiltonian. In a previous work~\cite{PhysRevB.95.085126} we have described a computationally
convenient scheme for extracting the SO coupling matrix elements for MLWFs. Here we extend the method to the computation of the spin-phonon
matrix elements. Our derived Hamiltonian can be readily used to compute spin-transport quantities, such as the spin relaxation length.

The paper is organized as follows. In the next section we introduce our computational approach and describe the 
specific implementation used. Then we present our results. We analyze first the symmetry of the various matrix
elements by considering the simple case of a linear atomic chain of Pb atoms. Then we move to the most complex case
of the durene crystal, a popular high-mobility organic semiconductor. Finally we conclude.

\section{Method}

Wannier functions, which form the basis functions of the proposed TB Hamiltonian, are essentially 
the weighted Fourier transforms of the Bloch states of a crystal. From a set of $N^\prime$ isolated Bloch 
states, $\{\ket{\psi_{m\mathbf{k}}}\}$, which for instance can be the Kohn-Sham (KS) eigenstates of 
a DFT calculation, one can obtain $N^\prime$ Wannier functions. The $n$-th Wannier ket centred at the 
lattice site $\mathbf{R}$, $\ket{w_{n\mathbf{R}}}$, is found from the prescription,
\begin{equation}\label{equ1}
\begin{split}
\ket{w_{n\mathbf{R}}}=\frac{V}{(2\pi)^3}\int_\mathrm{BZ}\left[\sum_{m=1}^N U_{mn}^{\mathbf{k}} \ket{\psi_{m\mathbf{k}}}\right]e^{-i\mathbf{k}.\mathbf{R}}d\mathbf{k}\:,
\end{split}
\end{equation}
where $V$ is the volume of the primitive cell, $\ket{\psi_{m\mathbf{k}}}$ is the $m$-th bloch vector, 
and the integration is performed over the first Brillouin zone (BZ). Here $U^{\mathbf{k}}$ is a unitary 
operator that mixes the Bloch states. In order to fix the gauge choice brought by $U^{\mathbf{k}}$ 
one minimizes the spread of a Wannier function, which is defined as
\begin{equation}\label{equ2}
\begin{split}
\Omega=\sum_n\left[\bra{w_{n\mathbf{0}}}r^2\ket{w_{n\mathbf{0}}}-|\bra{w_{n\mathbf{0}}}\mathbf{r}\ket{w_{n\mathbf{0}}}|^2\right]\:.
\end{split}
\end{equation}
Such choice defines the so-called MLWFs~\cite{PhysRevB.56.12847}. We use the code 
{\sc wannier90}~\cite{Mostofi2008685} for construction of such MLWFs.

Since the TB Hamiltonian operator, $\hat{H}$, depends on the ionic positions, the ionic motions give rise 
to changes in $\hat{H}$. In addition, since the MLWFs are constructed from the Bloch states, which 
themselves depend on the ionic coordinates, lattice vibrations result in a change of the MLWFs as well. 
Therefore, the change in the Hamiltonian matrix elements due to the ionic motion, namely the onsite energies 
and hopping integrals, originates from the combined action of 1) the change in $\hat{H}$ and 2) the change in 
the MLWFs basis. Hence, in the MLWFs TB picture the variation of the matrix element, $\varepsilon_{nm}$, 
due to an ionic displacement, is given by
\begin{equation}\label{ElPh_Coup_def_loc}
\Delta \varepsilon_{nm}=\braket{w_n^f|\hat{H}^f|w_m^f}-\braket{w_n^i|\hat{H}^i|w_m^i}\:,
\end{equation} 
where $w_m^i$ ($w_m^f$) and $\hat{H}^i$ ($\hat{H}^f$) are the initial (final) MLWF and the Hamiltonian 
operator, respectively. Eq.~(\ref{ElPh_Coup_def_loc}) describes the variation of an onsite energy or a hopping 
integral depending on whether $|w_m\rangle$ and $|w_n\rangle$ are located on the same site or at different 
sites. Since any general lattice vibration can be expanded as a linear combination of normal modes, one is 
typically interested in calculating $\Delta \varepsilon_{nm}$ due to vibrations along the normal mode coordinates. 
In order to quantify the rate of such change, we define the electron-phonon coupling parameter, 
$g^{\lambda}_{mn}$, for the $\lambda$-th phonon mode as the rate of change, $\Delta \varepsilon_{mn}$, of 
$\varepsilon_{mn}$ with respect to a displacement $\Delta Q_\lambda$ along such normal mode, namely
\begin{equation}\label{eq.ElPhCouplingDefinition}
g^{\lambda}_{mn}=\frac{\partial \varepsilon_{mn}}{\partial Q} \bigg\rvert_{Q \rightarrow Q+\Delta Q_\lambda}\:.
\end{equation}
Here $Q$ describes the system's geometry, so that $Q \rightarrow Q+\Delta Q_\lambda$ indicates that the 
partial derivative is to be taken with respect to the atomic displacement along the phonon eigenvector 
corresponding to the mode $\lambda$. 

This coupling constant is fundamentally different from that defined in a conventional TB formulation.
In that case the electron-phonon coupling is simply defined as
\begin{equation}\label{eq.Conventional_Coupling}
\alpha^\lambda_{nm}=\frac{\partial \left(\braket{\phi_n^i|\hat{H}^f-\hat{H}^i|\phi_m^i}\right)}{\partial Q} \bigg\rvert_{Q \rightarrow Q+\Delta Q_\lambda}\:,
\end{equation}
where $|\phi_n^i\rangle$ is the $n$-th basis function before the motion. Note that, at variance with 
Eq.~(\ref{ElPh_Coup_def_loc}), which takes into account both the changes in the operator and the basis 
set, in Eq.~(\ref{eq.Conventional_Coupling}) only the Hamiltonian operator is modified and the matrix 
element is evaluated with respect to the basis set corresponding to the equilibrium structure. For the 
remaining of this paper, unless stated otherwise, electron-phonon coupling will always denote the first description, i.e. the 
$g_{mn}^\lambda$s of Eq.~(\ref{eq.ElPhCouplingDefinition}). The effect of such coupling on charge transport 
has been the subject of many previous investigations.~\cite{PSSB:PSSB200743427,doi:10.1063/1.3033830,PhysRevB.95.035433,doi:10.1021/ct500390a}

As all matrix elements, also those associated to the SO coupling depend on the ionic coordinates. In a previous 
paper~\cite{PhysRevB.95.085126} we have described a method to calculate the SO matrix elements associated 
to the MLWFs basis, $\braket{w^{s_1}_{m\mathbf{R}}|\hat{V}_\mathrm{SO}|w^{s_2}_{n\mathbf{R'}}}$, from those computed 
over the spin-polarized Bloch states, $\braket{\psi^{s_1}_{m,\mathbf{k}}|\hat{V}_\mathrm{SO}|\psi^{s_2}_{n,\mathbf{k'}}}$
(the superscript denotes the magnetic spin quantum number). Note that here the MLWFs computed in absence of SO 
coupling are used as basis functions, since they span the entire relevant Hilbert space. The term 
$\braket{\psi^{s_1}_{m,\mathbf{k}}|\hat{V}_\mathrm{SO}|\psi^{s_2}_{n,\mathbf{k'}}}$ can be, in principle, calculated from 
any DFT implementation that incorporates SO coupling. Our choice is the {\sc siesta} code~\cite{0953-8984-14-11-302}, 
which uses an on-site approximation~\footnote{In the on-site approximation a SO coupling matrix element is neglected 
unless both orbitals and the SO pseudopotential are on the same atom.} for the 
SO coupling and gives the SO elements in terms of a set of localized atomic orbitals 
$\{\ket{\phi^{s}_{\mu,\mathbf{R}_l}}\}$~\cite{0953-8984-18-34-012}. Hence, the basic flowchart for such calculation follows 
the general prescription
\begin{align}
\braket{\phi^{s_1}_{\mu,\mathbf{R}_j}|\hat{V}_\mathrm{SO}|\phi^{s_2}_{\nu,\mathbf{R}_l}}\rightarrow \braket{\psi^{s_1}_{m,\mathbf{k}}|\hat{V}_\mathrm{SO}|\psi^{s_2}_{n,\mathbf{k'}}}\rightarrow \braket{w^{s_1}_{m\mathbf{R}}|\hat{V}_\mathrm{SO}|w^{s_2}_{n\mathbf{R'}}}\:,
\end{align}
namely from the SO matrix elements calculated for the {\sc siesta} local orbitals one computes those over
the Bloch functions and then the ones over the MLWFs.

Once the matrix elements $\braket{w^{s_1}_{m\mathbf{R}}|\hat{V}_\mathrm{SO}|w^{s_2}_{n\mathbf{R'}}}$ are known, 
it is possible to determine the spin-phonon coupling by following a prescription similar to that used for computing
the electron-phonon coupling in Eq.~(\ref{eq.ElPhCouplingDefinition}),
\begin{equation}\label{eq.SpPhCouplingDefinition}
g^{s_1s_2(\lambda)}_{m,n}=\frac{\partial \varepsilon_{\mathrm{(SO)}mn}^{s_1s_2}}{\partial Q} 
\bigg\rvert_{Q \rightarrow Q+\Delta Q_\lambda}\:,
\end{equation}
where $\varepsilon_{\mathrm{(SO)}mn}^{s_1s_2}$ is the SO matrix element between the MLWFs 
$\ket{w_m^{s_1}}$ and $\ket{w_n^{s_2}}$, $Q$ denotes the atomic positions and $\Delta{Q}$ refers 
to an infinitesimal displacement of the coordinates along the $\lambda$-th phonon mode. As noted 
earlier, a change in atomic coordinates results in a change in the MLWFs and such change must be 
taken into account when calculating the difference in the SO elements 
$\Delta \varepsilon_{\mathrm{(SO)}mn}^{s_1s_2}$. We use the same symbol $g$ to denote both the 
electron-phonon and the spin-phonon coupling, since they can be distinguished by the presence or 
absence of the spin indices.

In practice, when calculating both the electron-phonon and the spin-phonon coupling each atom $i$ 
in the unit cell is infinitesimally displaced by $\Delta Q_\lambda \mathbf{e^i_\lambda}$ along the direction 
of the corresponding phonon eigenvector, $\mathbf{e^i_\lambda}$. Then the electron-phonon (spin-phonon) 
coupling is calculated as $\Delta \varepsilon_{mn}/\Delta Q_\lambda$ 
($\Delta \varepsilon_{\mathrm{(SO)}mn}^{s_1s_2}/\Delta Q_\lambda$), i.e. from finite differences. If $\Delta Q_{\lambda}$ 
is too large, then the harmonic approximation, which is the basis of this approach, breaks down. In contrast, 
if $\Delta Q_{\lambda}$ is too small, then the quantity will have a significant numerical error. Hence, for any 
system studied, one must evaluate the coupling term for a range of $\Delta Q_\lambda$ and, from a plot of 
coupling terms vs $\Delta Q_\lambda$, choose the most suitable value of $\Delta Q_\lambda$. It is important 
to note that the coupling terms so defined have the dimension of energy/length. This is consistent with the 
semiclassical TB Hamiltonian used, for example in Ref.~\cite{PhysRevLett.96.086601}, for treating transport 
in organic crystals with significant dynamic disorder. However, various other definitions and dimensions for the 
electron-phonon coupling can be found in 
literature.~\cite{PhysRevB.85.115317,PhysRevB.82.035208,doi:10.1021/ct500390a,doi:10.1021/cr050140x}

\section{Results and Discussion}
\subsection{One Dimensional Pb Chain}

\begin{figure}
\centering
\includegraphics[width=0.50\textwidth]{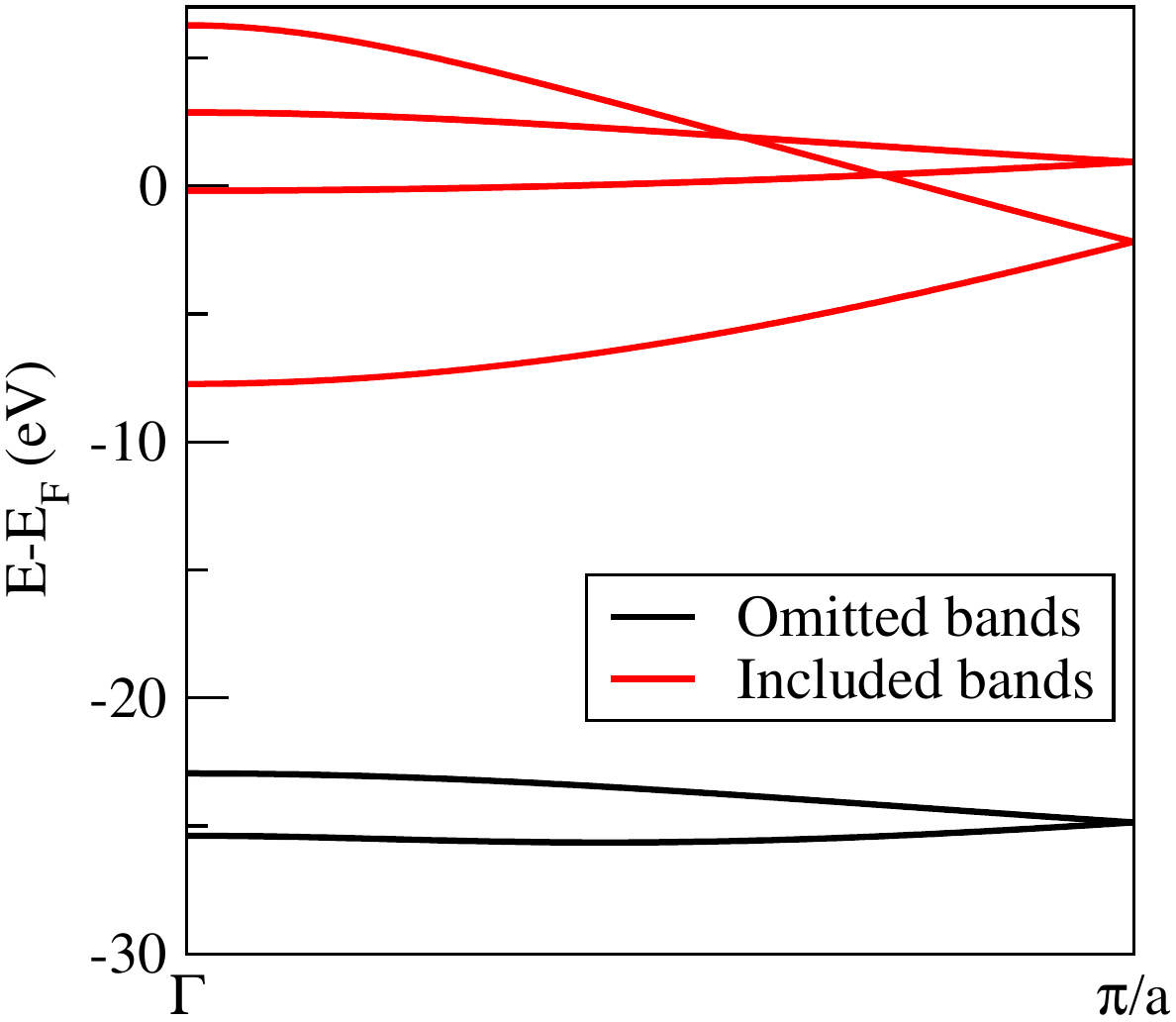}
\captionsetup{justification=justified, singlelinecheck=false}
  \caption{Band structure of a diatomic Pb chain calculated with a minimal basis set in {\sc siesta}. 
  The black and the red lines correspond to bands omitted from and included in the construction of 
  the MLWFs, respectively.}
\label{LeadBand}
\end{figure}
\begin{figure}
 \centering
  \includegraphics[width=0.50\textwidth]{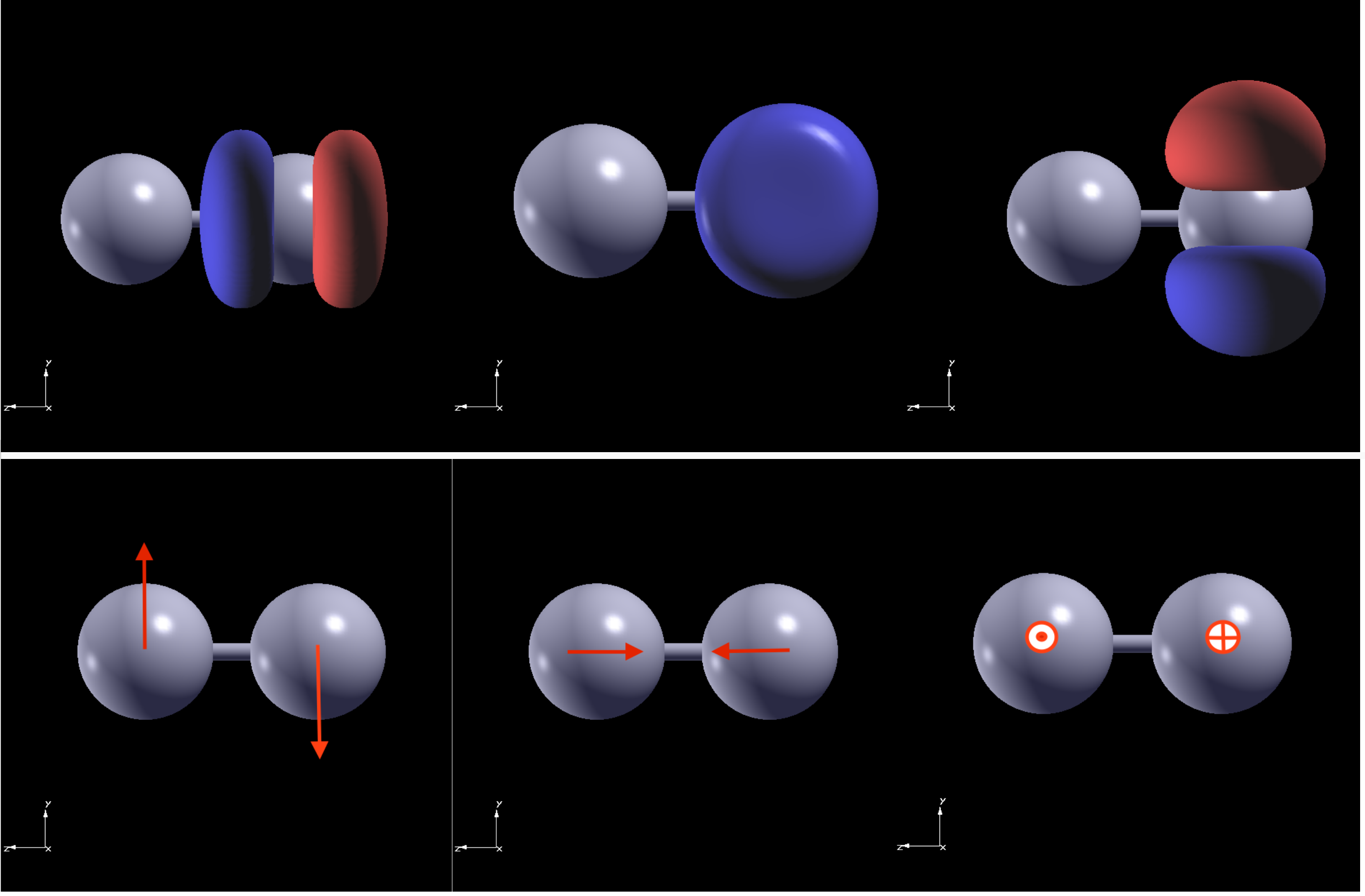}
  \captionsetup{justification=justified, singlelinecheck=false}
  \caption{The unit cell of the Pb chain containing two atoms. The figures in the top panel show isovalue plots 
  of the three MLWFs (from left to right: $\ket{w_{1,\mathbf{0}}}$, $\ket{w_{2,\mathbf{0}}}$ and 
  $\ket{w_{3,\mathbf{0}}}$) centred on the first atom. The bottom panels indicate the directions of the 
  atomic motion corresponding to the three phonon modes (mode 1, mode 2 and mode 3, from left to 
  right).}
  \label{fig:Pb_MLWF_Phonon}
 \end{figure}

A linear chain of Pb atoms with a diatomic unit cell has 6 phonon modes for each wave-vector, $\mathbf{q}$. 
For simplicity we restrict our calculations to the $\Gamma$-point, $\mathbf{q}=\mathbf{0}$, so that equivalent 
atoms in all unit cells have the same displacements with respect to their equilibrium positions. Since for the 
acoustic modes there is no relative displacement between the atoms of a unit cell, we are left with three 
optical modes of vibration as shown in the bottom panel of Fig.~\ref{fig:Pb_MLWF_Phonon}. 
The electronic band structure of a diatomic Pb chain calculated with a single-zeta basis functions is shown in 
Fig.~\ref{LeadBand}. Note that two of the bands marked in red are composed mostly of $p$ orbitals $\pi$-bonding 
and are doubly degenerate. Thus, as expected, the band structure contains 8 bands in total. The MLWFs are 
constructed by omitting the lowest two bands (mostly made of $s$-orbitals) and retaining the remaining 6 bands. 
This gives us six MLWFs per unit cell, three centred on each atom. For each of the three modes, we evaluate 
the coupling matrix elements between the MLWFs of the same unit cell for a range of $\Delta Q_{\lambda}$. 
By analysing these results we find that $\Delta Q_{\lambda}=0.03$ is an acceptable value for such fractional 
displacement.

The top panel of Fig.~\ref{fig:Pb_MLWF_Phonon} shows the MLWFs corresponding to the first atom of the 
unit cell at the equilibrium geometry. From this figure one can see that $\ket{w_{1,\mathbf{0}}}$, 
$\ket{w_{2,\mathbf{0}}}$ and $\ket{w_{3,\mathbf{0}}}$ closely resemble the $p$ orbitals of the first atom, 
which we can denote arbitrarily (the definition of the axes is arbitary) as $p_z$, $p_x$ and $p_y$, respectively.
By symmetry,  $\ket{w_{4,\mathbf{0}}}$, $\ket{w_{5,\mathbf{0}}}$, $\ket{w_{6,\mathbf{0}}}$ can be associated 
with the $p_z$, $p_x$ and $p_y$ orbitals located on the second atom. However, it is important to note that 
\textit{such similarity between the MLWFs and the orbital angular momentum eigenstates does not mean 
that they are equivalent}. In order to appreciate this point, note that
\begin{itemize}
\item $\braket{w_{i,\mathbf{0}}|w_{j,\mathbf{0}}}=0, \forall i \neq j$ but this is not necessarily true for 
$\braket{p_{m,1}|p_{n,2}}$, where $\ket{p_{m,1}}$ and $\ket{p_{n,2}}$ are orbital angular momentum 
eigenkets centred on the first and the second atom, respectively.
\item When an atom is displaced from its equilibrium position, the $p$ orbitals (e.g. the basis orbitals of 
{\sc siesta}) experience a rigid shift only, but do not change in shape. In contrast, the MLWFs change in 
shape along with being displaced.
\item Most importantly, in the on-site SO approximation used in {\sc siesta}, the hopping term for SO 
coupling, i.e. the SO matrix element between two orbitals located on two different atoms, is always zero. 
As for the on-site term, the SO matrix element between two orbitals of the same atom is independent of 
the position of the other atom. Thus, the spin-phonon matrix elements are always zero, when calculated
with the on-site SO approximation over the {\sc siesta} basis set. This is not the case for the MLWFs. 
Even when used in conjunction with an on-site SO approximation, the spin-phonon coupling is typically 
non-zero for a MLWF basis owing to the change in the basis functions upon ionic displacement.
\end{itemize} 

\begin{table}
\centering
\begin{tabular}{ccc}
Mode & Element & value(meV/\AA)\\
\hline
\hline
Mode 1 &$[w_3|w_4]$ & -0.85\\
\hline
\hline
Mode 2 &$[w_1|w_4]$ & 4.03\\
&$[w_2|w_5]$ & -1.51\\
&$[w_3|w_6]$ & -1.51\\
\hline
\hline
Mode 3 &$[w_2|w_4]$ & -0.85\\
\hline
\end{tabular}
\caption{The non-vanishing electron-phonon coupling matrix elements for the $\Gamma$-point phonon modes 
of the Pb chain with a diatomic unit cell. $[w_\mu|w_\nu]$ denotes the electron-phonon coupling matrix 
element between the MLWFs $\ket{w_\mu}$ and $\ket{w_\nu}$. One must keep in mind that the matrix 
elements are real and the remaining non-vanishing ones not reported in the table can be found from the 
relation $[w_\mu|w_\nu]=[w_\nu|w_\mu]$. See Fig.~\ref{fig:Pb_MLWF_Phonon} for a diagram of the 
modes and the MLWFs.}\label{table:LeadPhonon}
\end{table}

\begin{table}
\centering
\begin{tabular}{ccc}
Mode & Element & Value(meV/\AA)\\
\hline
\hline
Mode 1 &$[w_1^{\uparrow}|w_5^{\uparrow}]$ & (0.0,-0.07)\\
& $[w_2^{\uparrow}|w_4^{\uparrow}]$ & (0.0,0.07)\\
& $[w_2^{\uparrow}|w_6^{\downarrow}]$ & (-0.19,0.0)\\
& $[w_3^{\uparrow}|w_5^{\downarrow}]$ & (0.19,0.0)\\
\hline
\hline
Mode 2 & $[w_1^{\uparrow}|w_5^{\downarrow}]$ & (0.05,0.0)\\
& $[w_2^{\uparrow}|w_4^{\downarrow}]$ & (-0.05,0.0)\\
& $[w_1^{\uparrow}|w_6^{\downarrow}]$ & (0.0,-0.05)\\
& $[w_3^{\uparrow}|w_4^{\downarrow}]$ & (0.0,0.05)\\
\hline
\hline
Mode 3 &$[w_1^{\uparrow}|w_6^{\uparrow}]$ & (0.0,0.07)\\
& $[w_3^{\uparrow}|w_4^{\uparrow}]$ & (0.0,-0.07)\\
& $[w_2^{\uparrow}|w_6^{\downarrow}]$ & (0.0,-0.19)\\
& $[w_3^{\uparrow}|w_5^{\downarrow}]$ & (0.00,0.19)\\
\hline
\end{tabular}
\caption{Spin-phonon coupling matrix elements for the $\Gamma$-point phonon modes of 
the Pb chain with diatomic unit cell. $[w_\mu^{s_1}|w_\nu^{s_2}]$ denotes the complex 
spin-phonon coupling matrix element between the MLWFs $\ket{w_\mu^{s_1}}$ and $\ket{w_\nu^{s_2}}$. 
The remaining non-vanishing matrix elements can be found from the relations in Eq.~(\ref{eq.ID_SpPh_Pb}). 
The phonon modes and the MLWFs are shown in Fig.~\ref{fig:Pb_MLWF_Phonon}}\label{table:LeadSpinPhonon}.
\end{table}

Before calculating the spin-phonon coupling, let us take a brief look at the electron-phonon coupling matrix elements 
for the three phonon modes. The non-zero matrix elements are presented in Tab.~\ref{table:LeadPhonon} for each of the
normal modes. It is interesting to note that the change in overlap between the associated `$p$' orbitals due to the atomic 
displacements corresponding to the normal modes can be intuitively expected to have the same trend as the 
electron-phonon coupling matrix elements calculated with respect to the MLWFs (since the MLWFs closely resemble $p$ 
orbitals). For example, for an atomic motion along mode 3 (see Fig.~\ref{fig:Pb_MLWF_Phonon}), $\braket{p_{y,1}|p_{z,2}}$ 
must be zero, since $\ket{p_{z,2}}$ has always equal overlap with the positive and negative lobe of $\ket{p_{y,1}}$. Keeping 
in mind that modes 1, 2 and 3 correspond, respectively, to a motion in the $y$, $z$ and $x$ direction, one can easily show 
that
\begin{itemize}
\item $\Delta \braket{p_{z,1}|p_{z,2}}_{\rm mode:2} > \Delta \braket{p_{z,1}|p_{x,2}}_{\rm mode:3}$,
\item $\Delta \braket{p_{z,1}|p_{x,2}}_{\rm mode:3}$

$=\Delta \braket{p_{x,1}|p_{z,2}}_{\rm mode:3}$

$=\Delta \braket{p_{y,1}|p_{z,2}}_{\rm mode:1}$,
\item $\Delta \braket{p_{x,1}|p_{y,2}}_{\rm mode:1}$

$=\Delta \braket{p_{x,1}|p_{z,2}}_{\rm mode:2}$

$=\Delta \braket{p_{y,1}|p_{z,2}}_{\rm mode:3}$

$=0$
\end{itemize}   
where $\Delta$ denotes a change in the overlap of the orbitals due to their corresponding atomic motion.

Now we proceed to present our results for the spin-phonon coupling. At variance with the electron-phonon 
coupling matrix elements, the spin-phonon ones are not necessarily real valued. For each mode of the three 
modes, the inequivalent non-zero spin-phonon coupling matrix elements are tabulated in Tab.~\ref{table:LeadSpinPhonon}. 
We denote the spin-phonon matrix element between $\ket{w_\mu^{s_1}}$ and $\ket{w_\nu^{s_2}}$ as 
$[w_\mu^{s_1}|w_\nu^{s_2}]$. All other (equivalent) non-zero spin-phonon matrix elements can be found from 
those presented in Tab.~\ref{table:LeadSpinPhonon} by using the following relations
\begin{align*}\label{eq.ID_SpPh_Pb}
[w_\mu^{\uparrow}|w_\nu^{\downarrow}]& =-[w_\mu^{\downarrow}|w_\nu^{\uparrow}]^*,\\ 
[w_\mu^{\uparrow}|w_\nu^{\downarrow}]& =[w_\nu^{\downarrow}|w_\mu^{\uparrow}]^*,\\ 
\Im [w_\mu^{\uparrow}|w_\nu^{\uparrow}]& =-\Im [w_\mu^{\downarrow}|w_\nu^{\downarrow}]. \numberthis
\end{align*}
Also, from the symmetry of the MLWFs, it is easy to show that
\begin{align}
[w_1^{\uparrow}|w_5^{\uparrow}]_{\rm Mode 1}&{}=-[w_2^{\uparrow}|w_4^{\uparrow}]_{\rm Mode 1},\\
[w_1^{\uparrow}|w_6^{\uparrow}]_{\rm Mode 3}&{}=-[w_3^{\uparrow}|w_4^{\uparrow}]_{\rm Mode 3}.
\end{align}

We have noted that in the on-site approximation, the spin-phonon coupling (according to our definition) of the 
Pb chain should be zero, when calculated over the {\sc siesta} basis set. However, if such on-site approximation 
is relaxed, one will be able to determine a number of analytical expressions for these coupling elements in terms 
of the change in orbital overlaps. It is interesting to note that the analytical expressions calculated in this way share 
many qualitative similarities with those presented in Tab.~\ref{table:LeadSpinPhonon}. 
We summarize the findings of this section by noting that the spin-phonon couplings matrix elements corresponding 
to the two equivalent normal modes show the expected symmetry. We have also seen that the non-zero spin-phonon 
coupling matrix elements for mode 2 are, in general smaller than those for the symmetry-equivalent modes 1 and 3.

\subsection{Durene Crystal}

\begin{figure}
\includegraphics[width=0.6\textwidth]{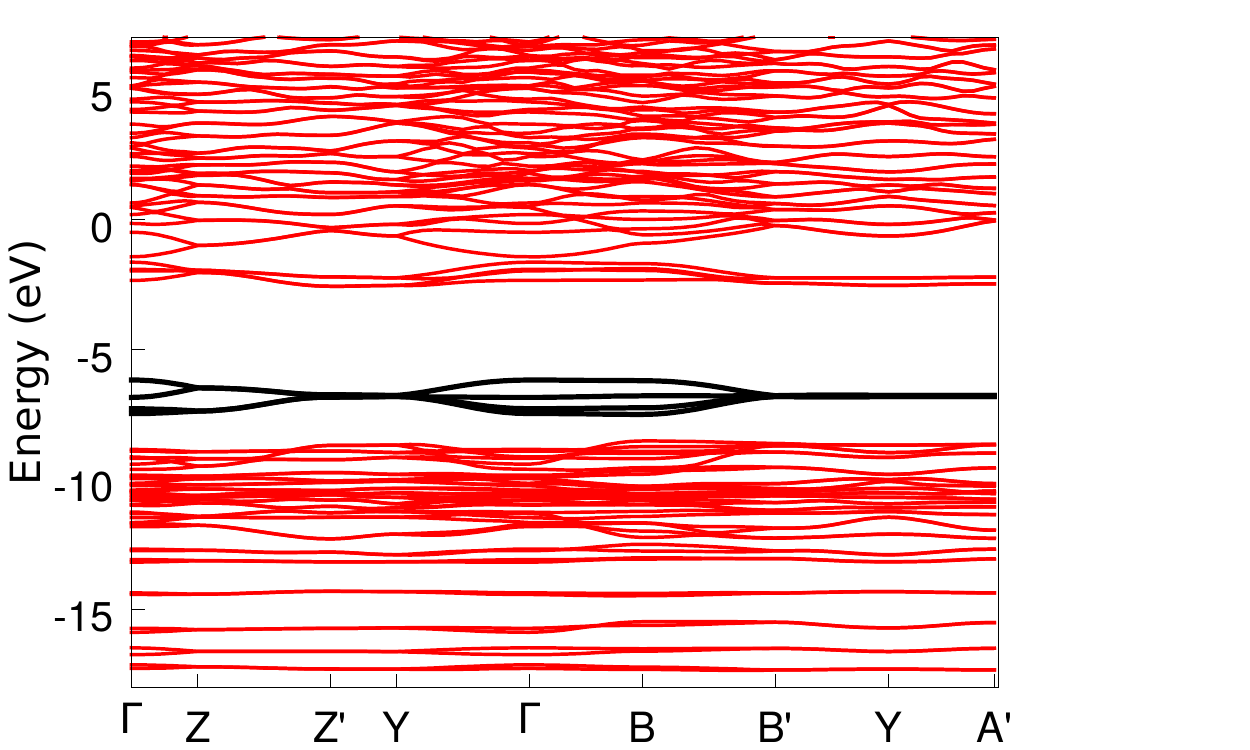}
\includegraphics[width=0.6\textwidth]{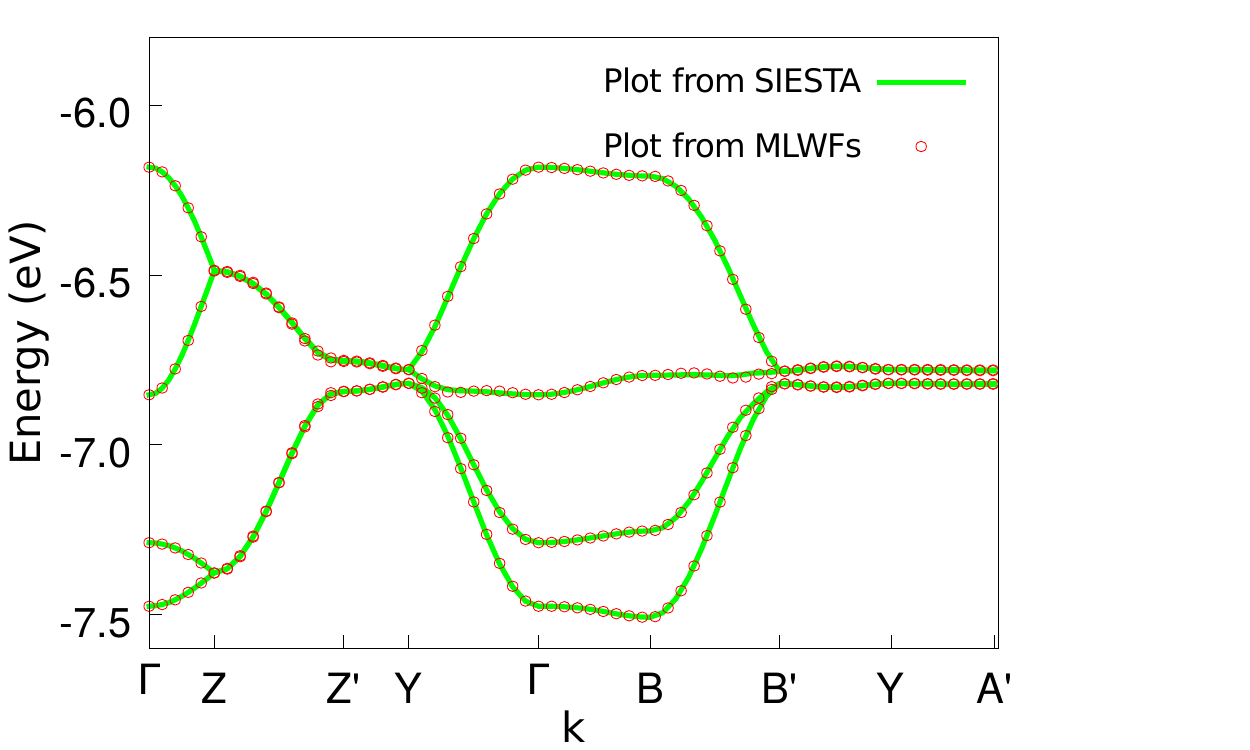}
\captionsetup{justification=justified,singlelinecheck=false}
\caption{Band structure of the durene crystal. Panel (a) shows all the occupied and many unoccupied bands. MLWFs 
are constructed from the 4 highest occupied bands, which are plotted in black. Panel (b) shows the magnified structure of 
these 4 bands plotted with {\sc siesta} (green line) and obtained from the MLWFs computed with {\sc wannier90} (red circle).}
\label{fig:DureneCrystalBandStructure}
\end{figure}

Finally we are in the position to discuss the spin-phonon coupling in a real organic crystal, namely in durene. In an electron-phonon or 
spin-phonon coupling calculation, one needs to make sure that the construction of the MLWFs converges to a global minimum, otherwise 
the various displaced geometries may correspond to different local minima resulting in the description of a different energy landscape. 
Typically, a MLWF calculation with dense $k$-mesh is likely to converge to a local minimum, while a calculation with coarse $k$-mesh 
has a higher probability of giving the global minimum ($\Gamma$-point calculation always converges to the global minimum). However, 
a coarse $k$-mesh translates in a small period for the Born-Von-Karman boundary conditions, i.e. a poorer description of the crystal. In 
our calculation, we use a $4\times 4\times4$ $k$-grid and construct the MLWFs from the top four valence bands. This enables the calculation 
to converge to a global minimum, identified by vanishing or negligible imaginary elements in the Hamiltonian matrix. In Fig.~\ref{fig:DureneCrystalBandStructure}(a) 
we show a plot of the durene bandstructure (within a large energy window) and in Fig.~\ref{fig:DureneCrystalBandStructure}(b), the 
bandstructure corresponding to the four bands used toconstruct MLWFs. These are plotted from the DFT {\sc siesta} eigenvalues and by 
diagonalizing the tight-binding Hamiltonian constructed over the MLWFs.

\begin{figure}
 \centering
  \fbox{\includegraphics[width=0.48\textwidth]{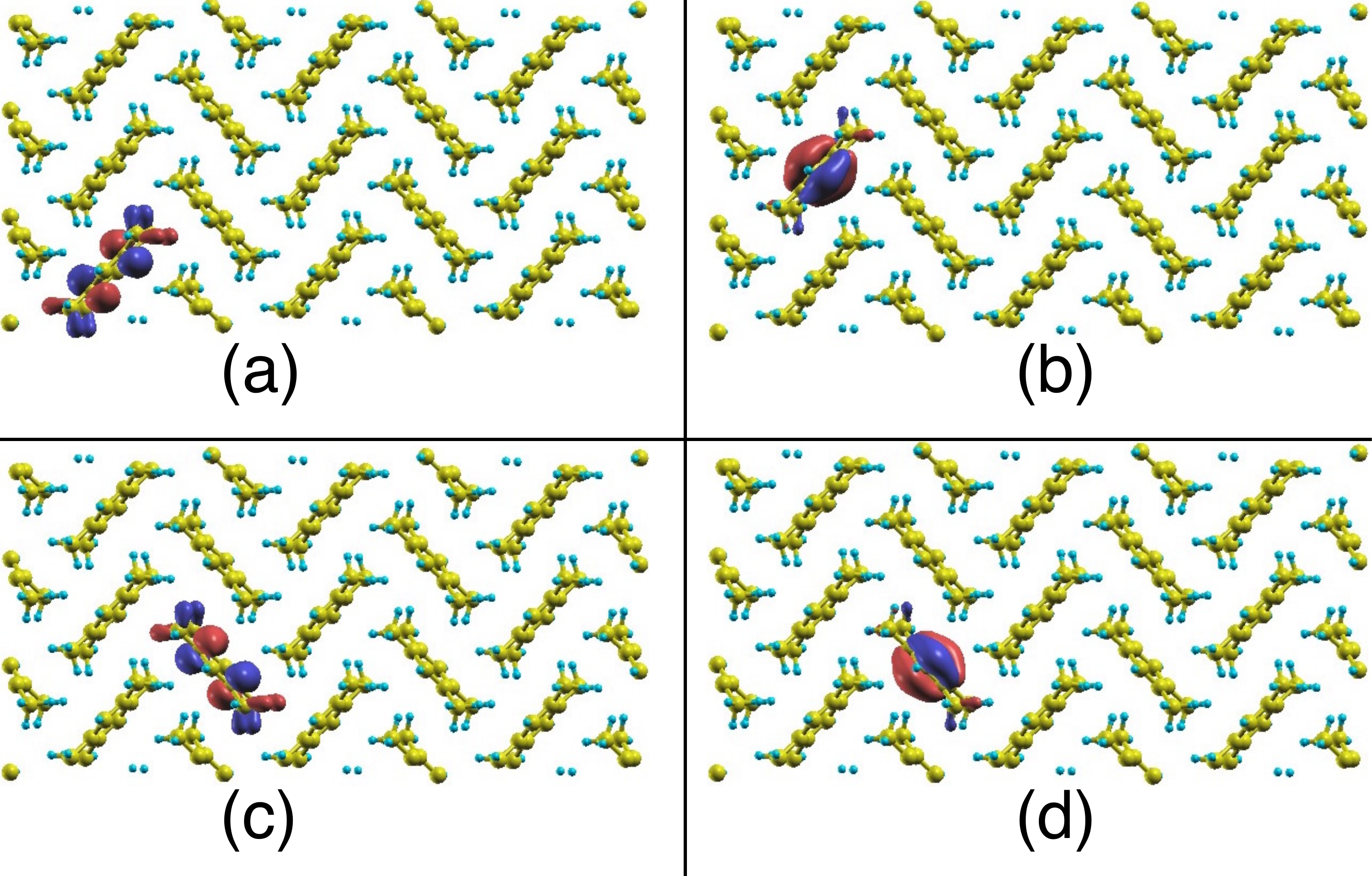}}
  \captionsetup{justification=justified, singlelinecheck=false}
  \caption{Isovalue plots for MLWFs of the four topmost valence bands of a durene crystal. Panels (a), (b), (c) and (d) correspond to 
  $\ket{w_{1,\mathbf{0}}}$, $\ket{w_{2,\mathbf{0}}}$, $\ket{w_{3,\mathbf{0}}}$ and $\ket{w_{4,\mathbf{0}}}$ respectively.}
  \label{fig:durene_MLWF_isovalue}
 \end{figure}
Since the unit cell of durene contains two molecules, the four valence bands give us four MLWFs per unit cell, so that each molecule 
has associated two MLWFs. In Fig.~\ref{fig:durene_MLWF_isovalue} we show an isovalue plot of the 4 MLWFs corresponding to 
$\mathbf{R}=\mathbf{0}$. We see that unlike $\ket{w_{3,\mathbf{0}}}$ and $\ket{w_{4,\mathbf{0}}}$, which are situated on the same 
molecule, $\ket{w_{1,\mathbf{0}}}$ and $\ket{w_{2,\mathbf{0}}}$ are on different but equivalent molecules displaced by a primitive lattice 
vector $\mathbf{a_2}$. Thus, $\ket{w_{1,\mathbf{0}}}$ and $\ket{w_{2,\mathbf{R'}}}$ are on the same molecule for $\mathbf{R'}=-\mathbf{a_2}$, 
where $\{\mathbf{a_1},\mathbf{a_2},\mathbf{a_3}\}$ is the set of primitive vectors. This means that for our tight-binding picture 
$\braket{w_{1,\mathbf{0}}|\hat{H}|w_{2,\mathbf{0}}}$ corresponds to a non-local (hopping) matrix element, whereas 
$\braket{w_{1,\mathbf{0}}|\hat{H}|w_{2,\mathbf{R'}}}$ is a local (on-site) energy term. In the following, we shall calculate the electron-phonon and 
spin-phonon coupling corresponding to various modes of the durene crystal and compare: 1) the relative contribution of the different modes,
2) for each mode, the relative contribution of the local and non-local terms.

Since the unit cell contains two molecules, each with 24 atoms (48 atoms in the unit cell), a $\Gamma$-point phonon calculation will give us 
144 modes, with 141 being non-trivial. Among these, 12 will be predominantly intermolecular modes (3 translational and 9 rotational modes, where
the molecules move rigidly with respect to each other) and the remaining ones will be of predominantly intramolecular nature. Here we shall 
consider only the phonon modes with an energy less than 75~meV, as the modes with higher energy are accessible only at high 
temperature~\cite{doi:10.1021/ct500390a}. Thus, we take into account 25 modes, of which the first 12 are intermolecular (these are lower in energy) 
and the rest are symmetry inequivalent intramolecular ones\footnote{The phonon spectrum is calculated with FHI-AIMS. Calculation courtesy: 
Dr. Carlo Motta}.

In order to compare the contributions of the different phonon modes and of the local (Holstein-type) and non-local (Peierls-type) contributions, we 
calculate the following effective electron-phonon coupling parameters
\begin{align}\label{Eq.EffectiveCoupling}
G^{\rm L}_\lambda&=\sum_{m,n} |g^\lambda_{mn}|^2\:,
\end{align}
where $m$ and $n$ are functions centred on same molecule, and
\begin{align}\label{Eq.EffectiveCoupling2}
G^{\rm N}_\lambda&=\sum_{m \neq n} |g^\lambda_{mn}|^2\:,
\end{align}
where $m$ and $n$ are on different molecules.

\begin{figure}
 \centering
  \includegraphics[width=0.5\textwidth]{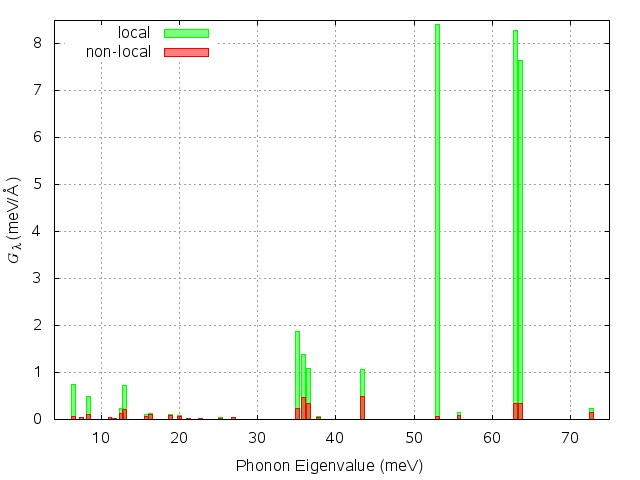}
  \captionsetup{justification=justified, singlelinecheck=false}
  \caption{Histogram of the effective electron-phonon coupling as a function of the phonons energy. The local and the non-local contributions are 
  denoted by green and red bars, respectively.}
  \label{fig:durene_ElPh}
 \end{figure}
Here the superscripts L and N stand for Local and Non-local, respectively. A crucial point to be noted for treating bulk crystals is that in 
{\sc wannier90}, the direct lattice points, where the MLWFs are calculated, are the lattice points of the Wigner-Seitz cell about the cell 
origin, $\mathbf{R}=\mathbf{0}$. Typically, one should expect the number of such lattice points to be the same as the number of $k$-points 
in reciprocal space. However, in a 3-D crystal it is possible to have lattice points, which are equidistant from the $\mathbf{R}=\mathbf{0}$ cell 
and (say) $n$ number of other cells. This means that such lattice point is shared by Wigner-Seitz cells of $n+1$ cells. In this case, this 
degenerate lattice point is taken into consideration by {\sc wannier90}, but a degeneracy weight of $1/(n+1)$ is associated with it. Consequently 
in further calculations (such as the band structure interpolation), its contribution carries a factor of $1/(n+1)$. Keeping this in mind, we multiply 
the contributions from the MLWFs of degenerate direct lattice points by their corresponding weighting factors. Fig.~\ref{fig:durene_ElPh} shows 
a histogram of the $G_\lambda$ terms as function of the phonons energy. It must be kept in mind that the coupling matrix elements are strongly 
dependent on the MLWFs. Therefore, constructing Wannier functions from a different set of Bloch states can in principle result in different values 
of $G_\lambda$. We see that in our case, most of the modes with high $G_\lambda(=G^{\rm L}_\lambda+G^{\rm N}_\lambda)$ are located at 
high phonon energies. Also, the electron-phonon couplings for modes with lower $G_\lambda$ are dominated by the non-local contributions, 
while those with higher $G_\lambda$ are dominated by local contributions.

Concerning the spin-phonon coupling, we can define spin-dependent $G_\lambda$ terms, namely the effective spin-phonon couplings, 
 \begin{align}\label{Eq.EffectiveSpPhCoupling}
G^{\rm L(s_1s_2)}_\lambda&=\sum_{m,n} |g^{s_1s_2(\lambda)}_{mn}|^2\:,
\end{align}
where $m$ and $n$ are on same molecule and
\begin{align}
G^{\rm N(s_1s_2)}_\lambda&=\sum_{m \neq n} |g^{s_1s_2(\lambda)}_{mn}|^2\:,
\end{align}
where $m$ and $n$ are on different molecules.
 
\begin{figure}
 \centering
  \includegraphics[width=0.5\textwidth]{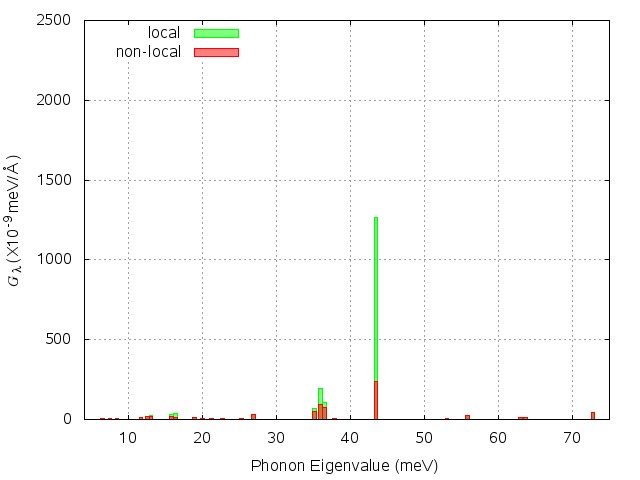}
  \includegraphics[width=0.5\textwidth]{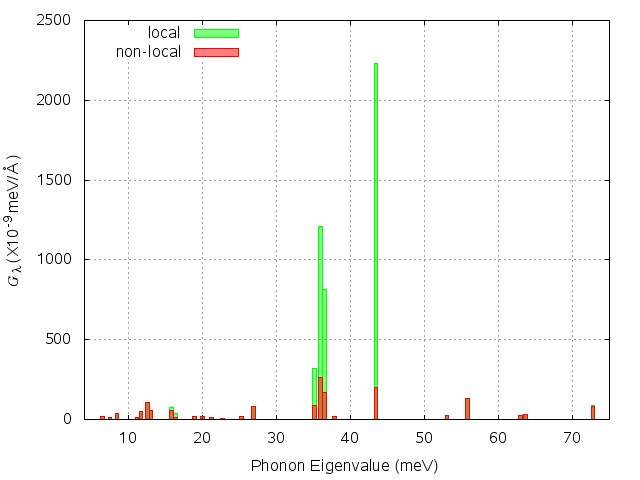}
  \captionsetup{justification=justified, singlelinecheck=false}
  \caption{Histogram plot of the effective spin-phonon coupling parameters, $G^{\rm(s_1s_2)}_\lambda$, as a function of the phonons energy. 
  The top panel corresponds to the ${\rm s_1=s_2}$ case (same spins), while the bottom one corresponds to ${\rm s_1\ne s_2}$ (different 
  spins). The local and the non-local contributions are denoted by green and red bars respectively.}
  \label{fig:durene_SpPh}
 \end{figure}
In Fig.~\ref{fig:durene_SpPh}, we plot these effective spin-phonon coupling terms, and we break down the local and non-local contributions. 
The top and the bottom panels correspond to the $(s_1=\uparrow,s_2=\uparrow)$ and $(s_1=\uparrow,s_2=\downarrow)$ case, respectively. 
As expected, the spin-phonon coupling terms are extremely small (about four orders of magnitude smaller than those of the Pb chain), owing 
to the small atomic masses in the crystal (the SOC is small). As in the case of the electron-phonon interaction, the effective spin-phonon coupling 
terms are dominated by non-local contributions for low $G_{\lambda}^{(s_1s_2)}=G^{\rm L(s_1s_2)}_\lambda+G^{\rm N(s_1s_2)}_\lambda$ and 
by local contributions for high $G_\lambda$. We also see that the spin-phonon coupling (for same spin, as well as for different spins) is very small 
for the first few modes, which represent intermolecular motions. This is fully consistent with the short-ranged nature of SO coupling. An important 
message emerging from these results is that phonon modes having high effective electron-phonon coupling do not necessarily have high effective 
spin-phonon coupling, and vice-versa. This means that the knowledge of the phonon spectrum says little {\it a priori} about the spin-phonon
coupling, so that any quantitative theory of spin relaxation cannot proceed unless a detail analysis along the lines outlined here is performed. 

In conclusion, we have discovered that both the electron-phonon and the spin-phonon coupling constants are, in general, dominated by the local 
modes, as expected by the short-range nature of the SOC. However, modes with very small effective coupling tend to have a larger relative 
contribution arising from non-local modes. No apparent correlation can be found between the effective coupling constants pertaining to various 
phonon modes for the electron-phonon coupling and those for the spin-phonon coupling.

\section{Conclusion}

Based on our previous work concerning the calculation of the SO matrix elements with respect to MLWFs basis sets, we have presented calculations 
of the spin-phonon coupling matrix elements of periodic systems. We note that, in order to be useful in a multiscale approach based on an effective 
Hamiltonian, the electron-phonon and the spin-phonon coupling are not to be calculated in terms of a fixed set of MLWFs. Instead, one must take into 
account the change in the MLWFs as a result of the ionic motions. The coupling matrix elements for a given phonon mode are calculated by displacing 
atoms from the ground state geometry along that phonon eigenvector and by taking finite differences. For phonon modes at the $\Gamma$-point, we 
have calculated the electron-phonon and spin-phonon coupling elements of a 1D chain of Pb atoms with two atoms per unit cell and of a bulk durene 
crystal. This latter is a widely-studied and well-known organic semiconductor. The spin-phonon coupling matrix elements of the Pb chain obey the 
expected symmetry relations. For durene we have observed that, in general, the spin-phonon coupling is dominated by local contributions (Holstein-modes), 
although, for phonon modes with a small net effective coupling, the non-local part seems to dominate. Our calculations of spin-phonon coupling matrix 
elements are expected to be valuable in the construction of a effective Hamiltonians to be used for computing transport-related quantities. This is 
particularly welcome in the case of organic crystals, where \textit{ab initio} computation of transport properties is a challenging task.

\section{Acknowledgements}
This work is supported by the European Research Council,
Quest project. Computational resources have been provided
by the supercomputer facilities at the Trinity Center for High
Performance Computing (TCHPC) and at the Irish Center
for High End Computing (ICHEC). We thank Carlo Motta for 
providing the $\Gamma$-point phonon eigenvalues and eigenvectors 
of durene crystal used in Ref.~\cite{doi:10.1021/ct500390a}.

%\bibliography{Bibi}

%merlin.mbs apsrev4-1.bst 2010-07-25 4.21a (PWD, AO, DPC) hacked
%Control: key (0)
%Control: author (8) initials jnrlst
%Control: editor formatted (1) identically to author
%Control: production of article title (-1) disabled
%Control: page (0) single
%Control: year (1) truncated
%Control: production of eprint (0) enabled
%

\end{document}